\documentclass[useAMS,usenatbib]{mnras}
\usepackage{times}
\usepackage{graphics,epsfig}
\usepackage{graphicx}
\usepackage{amssymb}
\usepackage{txfonts}

\usepackage{dcolumn}
\usepackage{bm}
\usepackage{natbib}
\usepackage{ulem}
\usepackage{color}

\newcommand{\kms}{km\,s$^{-1}$}

\newcommand{\nhi}{\ensuremath{N_{\rm HI}}}

\newcommand{\cm}{cm$^{-2}$}
\newcommand{\dv}{\Delta V_{\rm 20}}
\newcommand{\ddv}{\Delta V_{\rm 90}}
\newcommand{\ta}{\times 10^{7}}
\newcommand{\tb}{\times 10^{9}}
\newcommand{\tc}{\times 10^{14}}
\newcommand{\td}{\times 10^{20}}
\newcommand{\shidv}{\int S_{\rm HI} {\rm dV}}
\newcommand{\hi}{H{\sc i}}
\newcommand{\mhi}{M_{\rm HI}}
\newcommand{\msun}{M_{\odot}}
\newcommand{\hii}{H{\sc i} 21cm}

\title[The gas and stellar mass of low-$z$ DLAs]{The gas and stellar mass of low-redshift damped Lyman-$\alpha$ absorbers}

\author[Kanekar et al.]{Nissim Kanekar$^1$\thanks{E-mail: nkanekar@ncra.tifr.res.in}, 
Marcel Neeleman$^2$, J. Xavier Prochaska$^2$, and Tapasi Ghosh$^3$\\
$^1$National Centre for Radio Astrophysics, Tata Institute of Fundamental Research, 
Ganeshkhind, Pune - 411007, India\\
$^2$UCO/Lick Observatory, University of California -- Santa Cruz, Santa Cruz, CA 95064, USA\\
$^3$Arecibo Observatory, Arecibo, PR 00612, USA}

\begin{document}
\date{Accepted yyyy month dd. Received yyyy month dd; in original form yyyy month dd}

\maketitle
\label{firstpage}

\begin{abstract}
We report Hubble Space Telescope Cosmic Origins Spectrograph far-ultraviolet and Arecibo 
Telescope H{\sc i} 21cm spectroscopic studies of six damped and sub-damped Lyman-$\alpha$ absorbers (DLAs 
and sub-DLAs, respectively) at $z \lesssim 0.1$, that have yielded estimates of their H{\sc i} 
column density, metallicity and atomic gas mass. This significantly increases 
the number of DLAs with gas mass estimates, allowing the first comparison between the gas masses of 
DLAs and local galaxies. Including three absorbers from the literature, we obtain H{\sc i} 
masses $\approx (0.24 - 5.2) \times 10^9 \: {\rm M}_\odot$, lower than the knee of the local 
H{\sc i} mass function. This implies that massive galaxies do not dominate the absorption cross-section 
for low-$z$ DLAs. We use Sloan Digital Sky Survey photometry and spectroscopy to identify 
the likely hosts of four absorbers, obtaining low stellar masses, 
$\approx 10^7-10^{8.7} M_\odot$, in all cases, consistent with the hosts being dwarf galaxies. We obtain 
high H{\sc i} 21\,cm or CO emission line widths, $\Delta V_{20} \approx 100-290$~km~s$^{-1}$, 
and high gas fractions, $f_{\rm HI} \approx 5-100$, suggesting that the absorber hosts are gas-rich galaxies 
with low star formation efficiencies. However, the H{\sc i} 21\,cm velocity spreads ($\gtrsim 100$~km~s$^{-1}$) 
appear systematically larger than the velocity spreads in typical dwarf galaxies.


\end{abstract}


\section{Introduction} 
\label{sec:intro}

Absorption-selected galaxy samples, based on the presence of strong Lyman-$\alpha$ absorption 
in quasar spectra, are not biased towards high-luminosity objects and hence provide a 
view of ``normal'' galaxies at high redshifts. The highest \hi\ column density (\nhi)
systems, the damped and sub-damped Lyman-$\alpha$ absorbers (DLAs and sub-DLAs, respectively) have 
\nhi\ values similar to those in nearby gas-rich galaxies, and have hence been of much 
interest in studies of galaxy evolution \citep[e.g.][]{wolfe05}.

Absorption spectroscopy has yielded much information on DLAs, including their metallicities 
\citep[e.g.][]{prochaska03a,rafelski12}, gas temperatures \citep[e.g.][]{kanekar03,kanekar14}, and 
molecular fractions \citep[e.g.][]{ledoux03,noterdaeme08}. However, despite many searches, the galaxy 
counterparts of only a dozen DLAs and sub-DLAs, mostly targetted due to an atypically high 
metallicity, have so far been detected in optical/ultraviolet emission at $z \gtrsim 2$ 
\citep[e.g.][]{fynbo11,fynbo13}. Typical high-$z$ DLAs appear to have low in-situ star 
formation rates (SFRs), $\lesssim 0.3 \; \msun$~yr$^{-1}$ \citep{fumagalli15}. The 
situation is somewhat better at low redshifts, $z < 1$, with estimates of the SFR, stellar mass, 
etc. available for $\approx 25$ absorbers \citep[e.g.][]{peroux12}.

Our knowledge of the gas content of the absorbers is even worse than that of the stellar 
content. The radio \hii\ hyperfine and CO rotational transitions are the main probes of atomic 
and molecular gas in nearby galaxies. Unfortunately, few DLAs are known at low redshifts, $z \lesssim 0.2$, 
where the weak \hii\ line is detectable with today's radio telescopes. \hii\ emission has hence 
only been detected in one DLA, at $z \approx 0.009$ towards SBS~1543+593 \citep{bowen01b,chengalur02}, 
and one sub-DLA, at $z \approx 0.006$ towards PG~1216+069 \citep{briggs06,chengalur15}, with a 
few non-detections yielding limits on the \hi\ mass \citep{mazumdar14}. In the case of 
molecular gas, there is so far only a single CO detection, at $z=0.101$ towards PKS~0439$-$433 
\citep{neeleman16b}. And, while the recent detection of C{\sc ii}-158$\mu$m 
emission in two $z \approx 4$ DLAs \citep{neeleman17} provides an exciting new tool to identify
high-$z$ DLA host galaxies, this transition does not provide information on the gas 
content of the absorbers.

A detailed comparison between the stellar and gas properties of absorption-selected galaxies
requires a large absorber sample at low redshifts, $z \lesssim 0.2$. The excellent far-ultraviolet 
(FUV) sensitivity of the Cosmic Origins Spectrograph (COS) onboard the Hubble Space Telescope (HST) 
now allows the detection of Lyman-$\alpha$ absorption at very low redshifts. We have hence analysed 
the HST data archive \citep{neeleman16}, to identify low-$z$ absorbers suitable for follow-up 
studies to characterize the host galaxies. We have now used the Arecibo Telescope to carry out a 
search for \hii\ emission from a set of low-$z$ absorbers identified in our survey.  In this 
{\it Letter}, we present the $N_{\rm HI}$ values, metallicities, and atomic gas and stellar 
masses for six systems at $z < 0.1$.

\setcounter{table}{0}
\begin{table*}
\begin{center}
\caption{Column densities and abundances from the HST-COS spectroscopy.
\label{tab:data1} }
\begin{tabular}{|c|c|c|c|c|c|c|c|c|}
\hline 
QSO 	   & $z_{\rm DLA}$ & $\nhi$	     & N(O{\sc i})     & N(Si{\sc ii})   & N(S{\sc ii})    & N(Fe{\sc ii})     & [M/H]                   & M$^a$                        \\
           &               & $\td$~\cm       & $\tc$~\cm      & $\tc$~\cm      & $\tc$~\cm      & $\tc$~\cm        & 		              &                              \\
\hline                                                                                                                                       
J0930+2845 & $0.0228$      & $5.6 \pm 1.2$   & $> 13.8$        & $> 1.0$         & $< 13.5$        & $< 4.4$           & [$-2.26$ , $-0.77$]$^b$ & S{\sc ii}, Si{\sc ii}    \\ 
J0951+3307 & $0.0054$      & $10.0 \pm 2.5$  & $> 17.8$        & $> 7.9$         & $14.5 \pm 3.0$ 	& $> 6.9$      & $-0.99 \pm 0.14$        & S{\sc ii}                    \\ 
J1415+1634 & $0.0077$      & $0.5 \pm 0.1$   & $3.02 \pm 0.28$ & $0.35 \pm 0.02$ & $<1.7$          & $< 0.62$          & $-1.91 \pm 0.11$$^c$    & O{\sc i}                     \\ 
J1512+0128 & $0.0295$      & $2.5 \pm 0.6$   & $> 25.1$        & $4.27 \pm 0.88$ & $<17.4$         & $< 2.7$           & $-1.29 \pm 0.13$        & Si{\sc ii}                   \\ 
J1553+3548 & $0.0829$      & $0.5 \pm 0.1$   & $> 3.1$         & $1.32 \pm 0.09$ & $<3.1$          & $1.15 \pm 0.19$   & $-1.35 \pm 0.16$        & Si{\sc ii}                   \\ 
J1619+3342 & $0.0963$      & $4.0 \pm 1.4$   & $> 6.3$         & $>0.91$         & $4.57 \pm 0.53$ & $1.00 \pm 0.09$   & $-1.09 \pm 0.16$        & S{\sc ii}                    \\ 
\hline
\end{tabular}
\end{center}
\begin{center}
$^a$The element used in the metallicity estimate of the previous column. $^b$The allowed metallicity 
range; see main text for discussion. $^c$Ionization corrections have not been included, but are expected to be small, $\lesssim 0.3$~dex \citep[e.g.][]{battisti12}.\\
\end{center}
\end{table*}

\section{Observations, data analysis and spectra}
\label{sec:data}

\begin{figure*}
\centering
\includegraphics[scale=0.8]{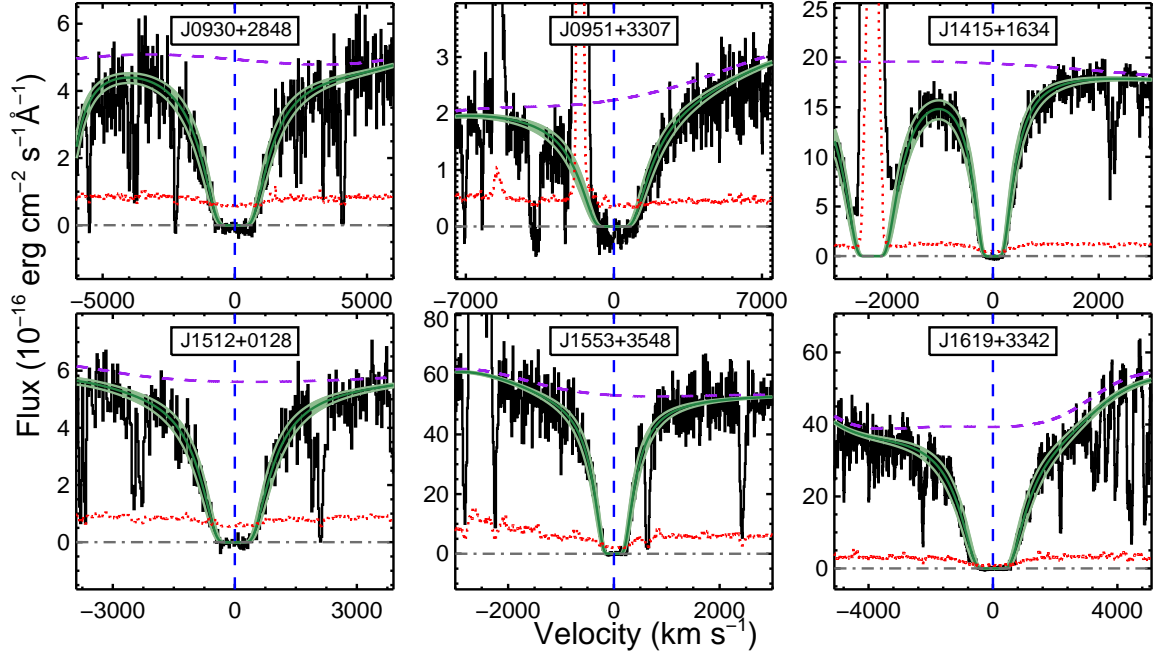}
\caption{The HST-COS Lyman-$\alpha$ profiles and our Voigt profile fits for the six absorbers
of our sample. The dashed and dotted curves show the quasar continuum and the error spectrum, respectively.
\label{fig:lya}}
\end{figure*}

\subsection{The HST observations}
\label{sec:sample}

The details of our analysis of the HST archival data on quasars observed with COS, the Space 
Telescope Imaging Spectrograph, or the Faint Object Spectrograph, are presented in 
\citet{neeleman16}. Standard pipelines were used to produce the final spectrum for each 
quasar. The search for Lyman-$\alpha$ absorption followed the approach of \citet{prochaska05}, with minor 
modifications \citep[see][]{neeleman16}.  For each absorber, \nhi\ was estimated using a 
custom IDL Voigt-profile fitting program, simultaneously fitting both the Voigt profile and the 
quasar continuum. The metal column densities were derived using the apparent optical depth method 
\citep{savage91}, and then used to infer the gas metallicity \citep{rafelski12}.

Our search yielded 15 DLAs and sub-DLAs at $z \lesssim 0.2$; note that this is {\it not} an unbiased 
sample (see Section 4).
We focus here on the six absorbers for which we were able to obtain \hii\ spectroscopy. Four 
systems are DLAs, with \nhi~$\geq 2 \times 10^{20}$~\cm, and two are sub-DLAs, both with 
\nhi~$=5 \times 10^{19}$~\cm. Their Lyman-$\alpha$ absorption profiles and the Voigt profile fits 
to estimate the \hi\ column density are shown in Fig.~\ref{fig:lya}, and their
redshifts, \hi\ and metal column densities, and metallicities are listed in Table~\ref{tab:data1}.


\subsection{The Arecibo observations}
\label{sec:arecibo}

\begin{figure*}
\centering
\includegraphics[scale=0.25]{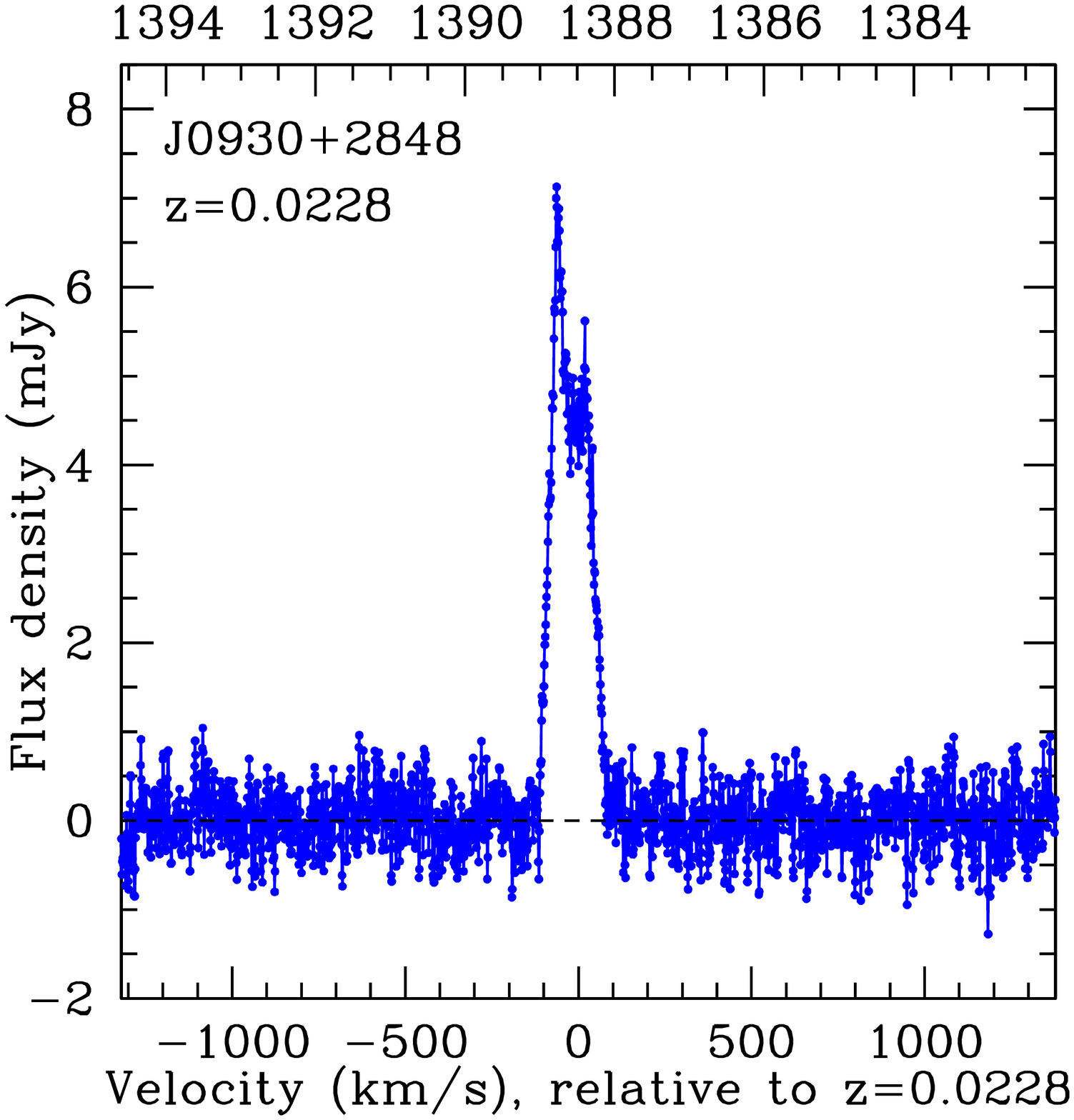}
\includegraphics[scale=0.25]{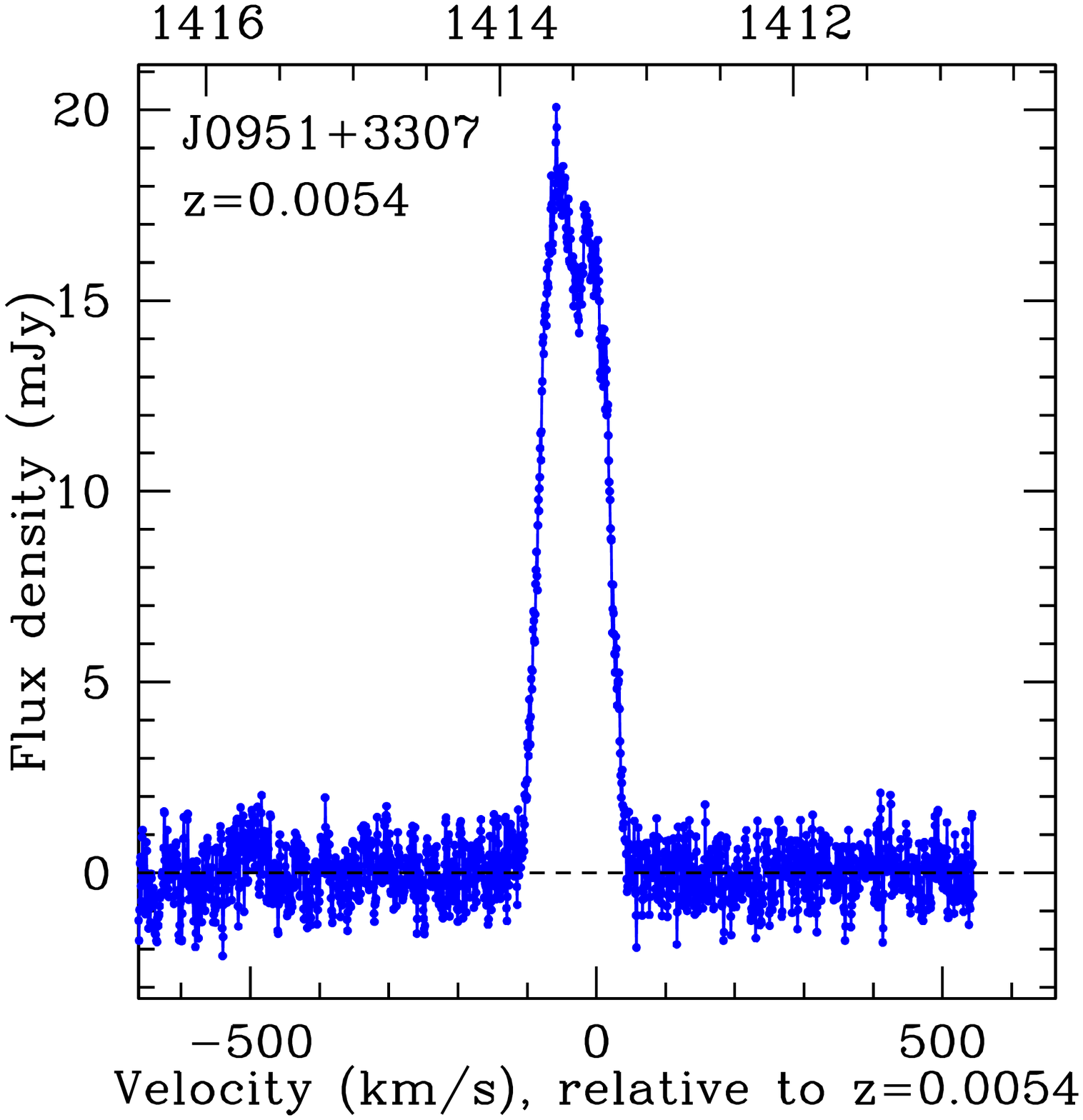}
\includegraphics[scale=0.25]{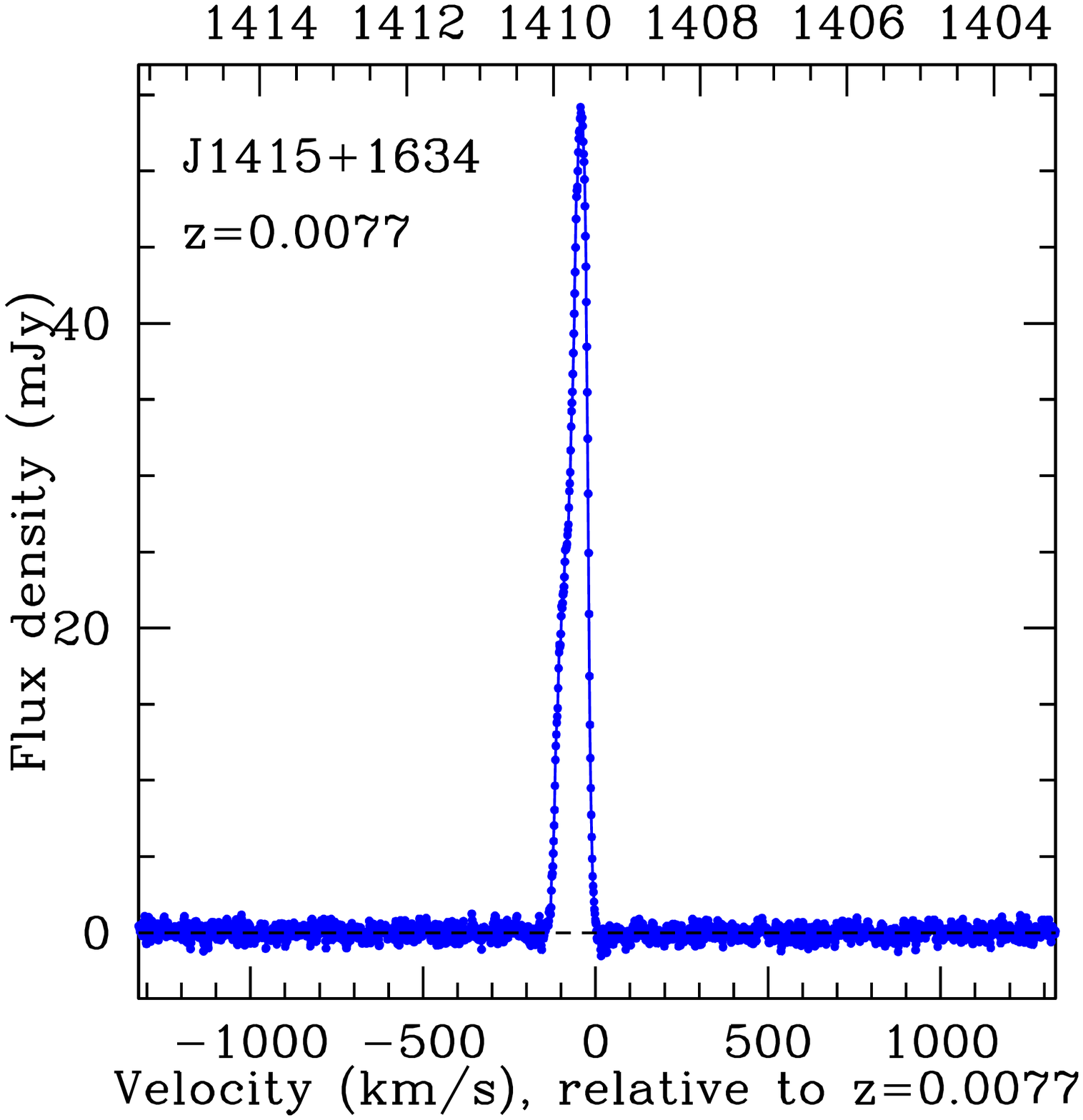}
\includegraphics[scale=0.25]{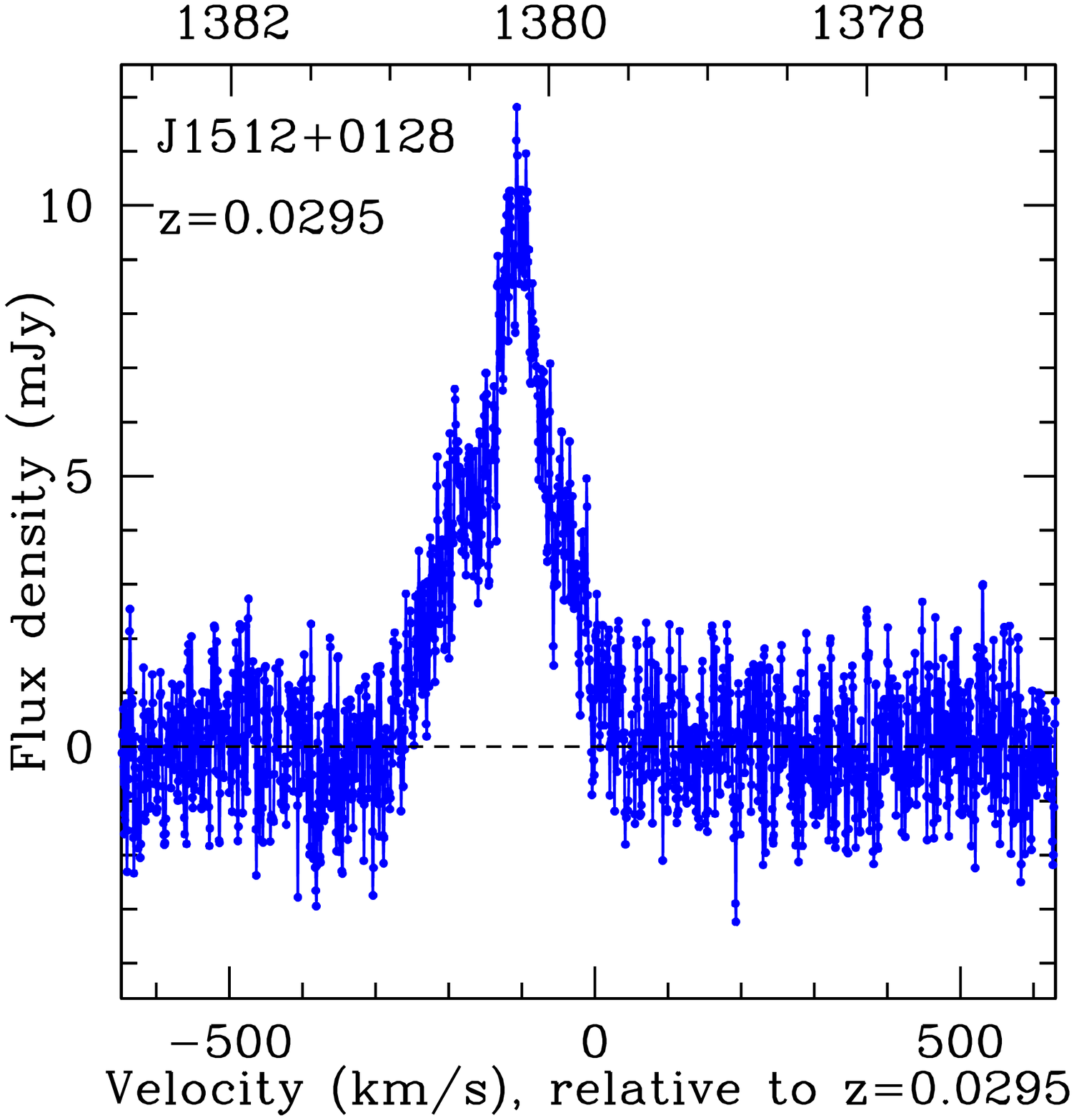}
\includegraphics[scale=0.25]{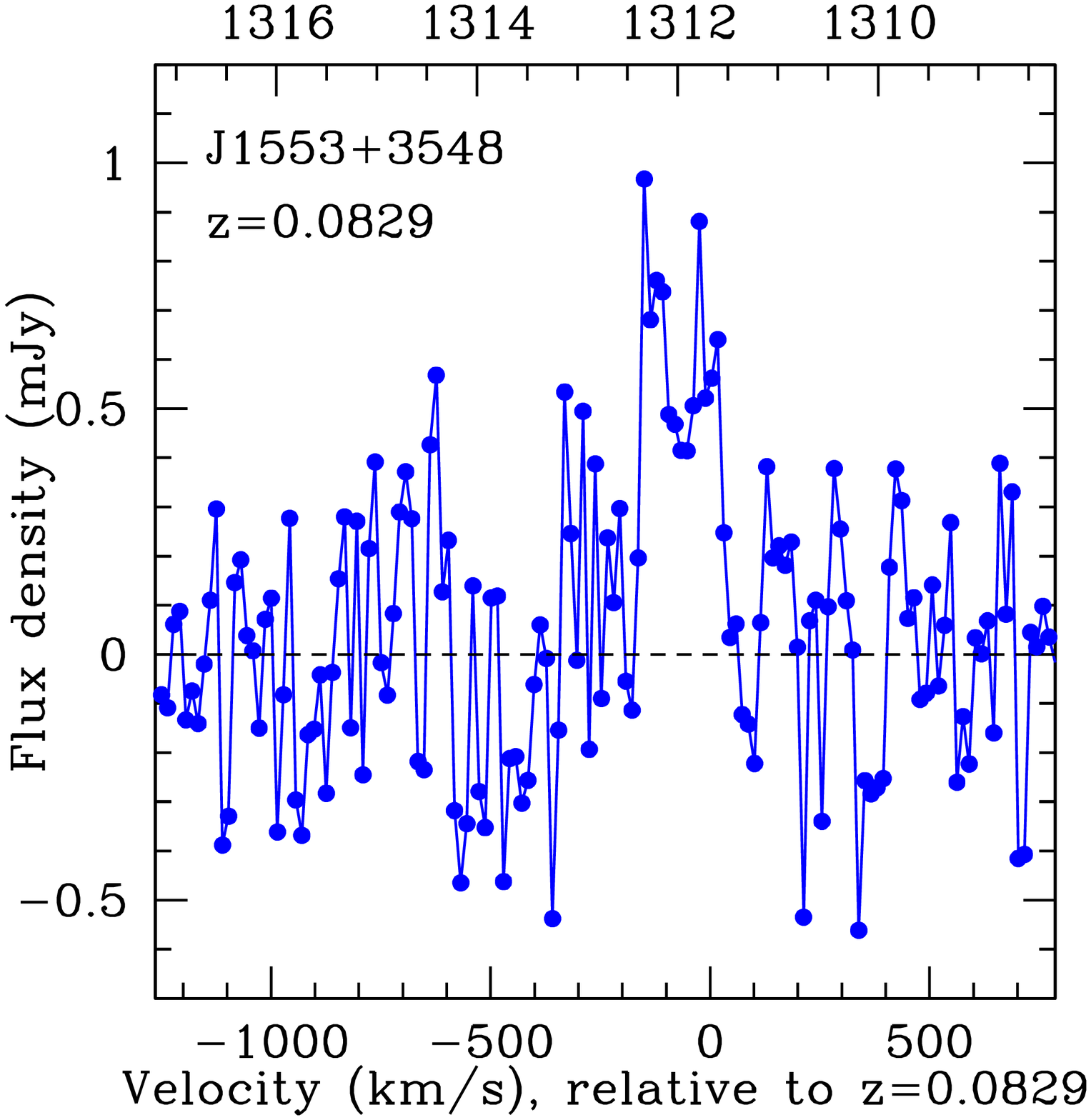}
\includegraphics[scale=0.25]{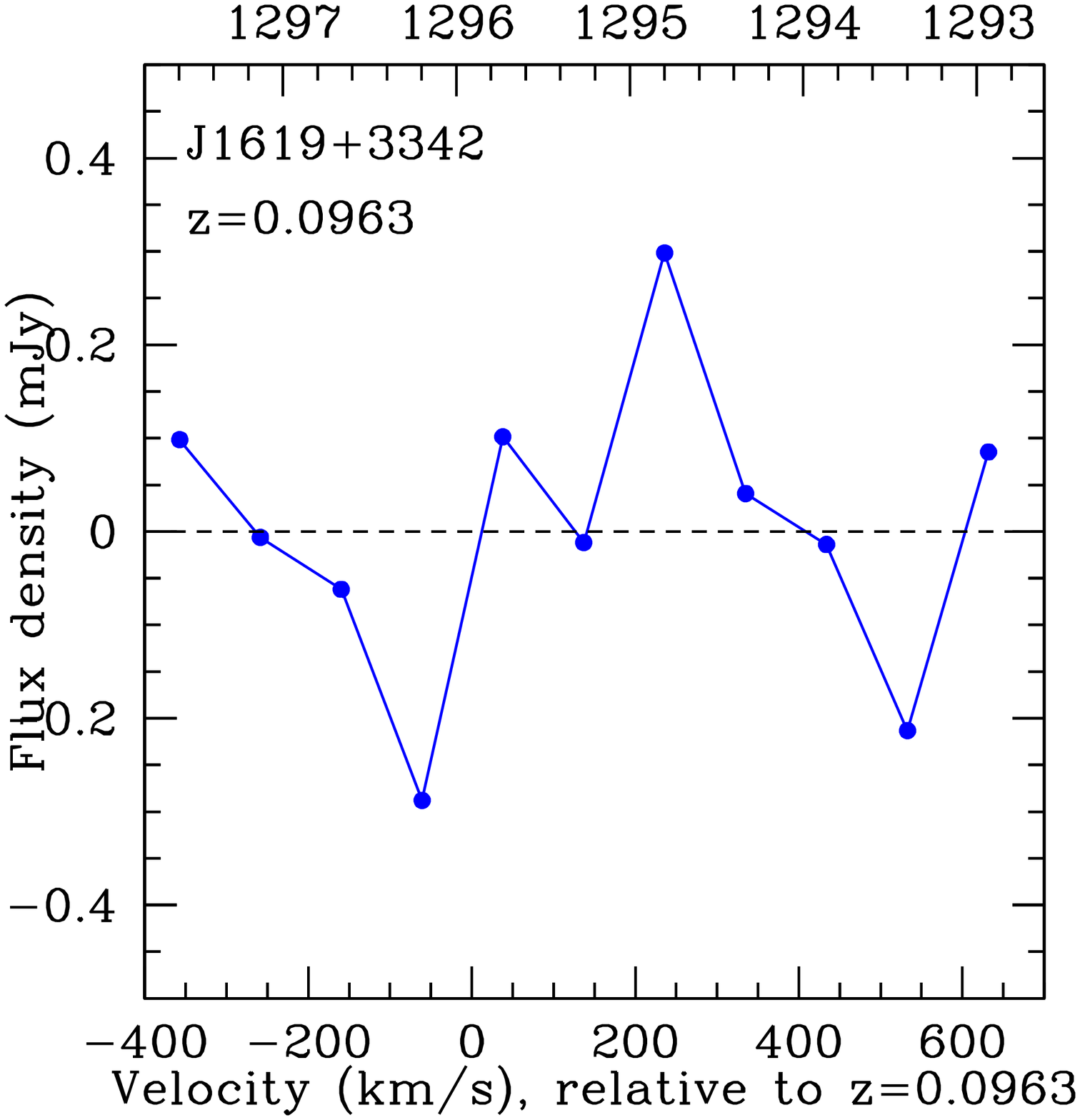}
\caption{The Arecibo \hii\ emission spectra from the six absorbers of our sample. 
}
\label{fig:hi}
\end{figure*}

We used the Arecibo L-Band-wide receiver over April--July~2015 and May~2016 in proposal A2940 
(PI: Kanekar) to search for redshifted \hii\ emission from 11 DLAs and sub-DLAs at $z \lesssim 0.2$,
observable with the Arecibo Telescope \citep[e.g.][]{meiring11,neeleman16}. Observations of five 
targets were affected by radio frequency interference (RFI); it was not possible to obtain clean spectra 
for these systems, which will not be discussed further. Bandwidths of 6.25~MHz, 12.5~MHz, 25~MHz and 
50~MHz were simultaneously used for the observations, centred at the expected redshifted \hii\ 
line frequency, and sub-divided into 2048 or 8192 channels, with the WAPP backend. Position switching 
(On/Off) was used for bandpass calibration, while the flux density scale was calibrated using a 
noise diode. The total on-source time was $\approx 1-3$~hours per source.

All the Arecibo data were analysed in IDL, using standard procedures. While the data on all six 
targets were relatively free of RFI, detailed flagging was necessary for the two systems at 
$z \approx 0.1$ (towards J1553+3548 and J1619+3342) to obtain clean spectra. The root-mean-square 
(RMS) noise values on the final spectra range from $0.33 - 0.68$~mJy per 12.2~kHz channel.

The final \hii\ spectra for the six DLAs and sub-DLAs are shown in Fig.~\ref{fig:hi}, in order
of increasing right ascension. We obtained five detections of \hii\ emission (with $> 5\sigma$ 
significance), out to $z = 0.0829$. For four of the six systems, the spectra are presented after 
Hanning-smoothing to, and resampling at, a resolution of 12.2~kHz ($\approx 2.6$~\kms\ at the respective
line frequencies). For the fifth detection, at $z=0.0829$ towards J1553+3548, we further boxcar-smoothed 
the spectrum by 5 channels; the spectrum of this source in Fig.~\ref{fig:hi} is at a resolution of $61$~kHz 
($\approx 13.9$~\kms\ at the redshifted \hii\ line frequency). Finally, in the case of the sole 
non-detection, for the $z = 0.0963$ DLA towards J1619+3342, the spectrum shown in Fig.~\ref{fig:hi} 
has been box-car smoothed to, and resampled at, a resolution of $100$~\kms.  Table~\ref{tab:data2} 
provides details of the results of the \hii\ observations.

\section{Results}

\begin{table*}
\begin{center}
\caption{Results from the SDSS photometry, and the HST-COS and Arecibo spectroscopy.
\label{tab:data2} }
\begin{tabular}{|c|c|c|c|c|c|c|c|c|c|}
\hline 
QSO 	   & $z_{\rm DLA}$ & $\nhi$	     & [M/H]                   & M$_*$	         &  $\shidv$$^\ast$  & $\mhi$$^a$$^\ast$ & $\dv^b$ & $\ddv^c$ & $f_{\rm HI}^d$ \\
	   & 	 	   & $\td$~\cm       & 		               & $\ta \; \msun$  &  Jy~\kms\	     & $\tb \; \msun$    &  \kms   & \kms     &   \\
\hline                                                                 
J0930+2845 & $0.0228$ 	   & $5.6 \pm 1.2$   & [$-2.26$ , $-0.77$]     & $6.9^{+4.6}_{-0.5}$           & $0.729 \pm 0.073$ & $1.75 \pm 0.18$   & $165$   & $153$    & $25$      \\
J0951+3307 & $0.0054$ 	   & $10.0 \pm 2.5$  & $-0.99 \pm 0.14$        & $4.0$           & $1.83 \pm 0.18$   & $0.237 \pm 0.024$ & $130$   & $94$     & $6$     \\
J1415+1634 & $0.0077$ 	   & $0.5 \pm 0.1$   & $-1.91 \pm 0.11$        & $1.0 - 2.6$     & $3.53 \pm 0.35$ & $0.965 \pm 0.097$   & $100$   & $118$    & $37-97$ \\
J1512+0128 & $0.0295$ 	   & $2.5 \pm 0.6$   & $-1.29 \pm 0.13$        & $-$             & $1.27 \pm 0.13$   & $5.15 \pm 0.52$   & $230$   & $267$    & $-$     \\
J1553+3548 & $0.0829$ 	   & $0.5 \pm 0.1$   & $-1.35 \pm 0.16$        & $23.4^{+34.6}_{-3.9}$          & $0.120 \pm 0.016$ & $3.93 \pm 0.61$   & $270$   & $74$     & $17$ 	  \\
J1619+3342 & $0.0963$ 	   & $4.0 \pm 1.4$   & $-1.09 \pm 0.16$        & $-$             & $< 0.014^e$       & $< 1.9^e$	 & $-$     & $22$     & $-$     \\
\hline
\multicolumn{10}{|c|}{Measurements from the literature}\\
\hline
PG~1216+069    & $0.0063$  & $0.21 \pm 0.01$ & $-1.60 \pm 0.10^f$      & $-$	         & $0.178^h$	     & $0.032^h$         & $100$   & $120$    & $-$ \\
SBS~1543+543   & $0.0096$  & $2.6 \pm 0.2$   & $-0.41 \pm 0.06^f$      & $4.5^g$         & $ 3.9^h$          & $1.5^h$  	 & $75$    & $128$    & $33$ \\
PKS~0439$-$433 & $0.1011$  & $0.43 \pm 0.03$ & $+0.28 \pm 0.08^f$      & $1023 \pm 47^g$ & $< 0.064^{e,h}$     & $< 3.1^{e,h}$   & $290^i$ & $109$  & $< 0.3$\\
\hline 
\end{tabular}
\end{center}
\begin{center}
$^\ast$The quoted errors are the sums in quadrature of 
the measurement errors and $\approx 10$\% systematic errors on the flux density scale.\\
$^a$The \hi\ masses assume a $\Lambda$-cold-dark-matter cosmology, with $H_0 = 67.8$~\kms~Mpc$^{-1}$, 
$\Omega_m = 0.308$, and $\Omega_\Lambda = 0.692$ \citep{planck16}.\\
$^b$$\dv$ is the velocity width between 20\% points of the peak \hii\ flux density.\\
$^c$$\ddv$ is the velocity width containing 90\% of the equivalent width of the low-ionization metal absorption lines. \\
$^d$The gas fraction, $f_{\rm HI}$, is defined here as the ratio of the \hi\ mass to the stellar mass \citep[][]{huang12}.\\
$^e$These are $3\sigma$ limits on the integrated \hii\ line flux density and the \hi\ mass, assuming a Gaussian 
line profile with an FWHM of $100$~\kms. \\
References for literature absorbers: $^f$Metallicity: \citet{dutta15}, \citet{tripp05}, \citet{bowen05};
$^g$Stellar mass: \citet{christensen14}, \citet{rosenberg06}; $^h$\hii\ data: \citet{bowen01b}, \citet{kanekar01e}, \citet{chengalur15}.\\
$^i$The value of $\dv$ for the $z=0.1010$ sub-DLA towards PKS~0439$-$433 is for the CO~J=1-0 line \citep{neeleman16b}.
\end{center}
\end{table*}

{\bf J0930+2845} $\bm{(z = 0.0228)}$:~The $z = 0.0228$ DLA towards J0930+2845 has 
\nhi~$=(5.6 \pm 1.2) \times 10^{20}$~\cm. Unfortunately, most of the metal lines 
covered by the HST-COS spectrum for this DLA are either saturated or not detected: we obtain 
log[N(Si{\sc ii})/\cm]$> 14.00$,
log[N(O{\sc i})/\cm]~$>15.25$, and log[N(S{\sc ii})/\cm]~$< 15.13$. Combining the Si{\sc ii} 
lower limit and the S{\sc ii} upper limit yields the metallicity range [M/H]~$=[-2.26,-0.77]$. 
The \hi\ mass of the galaxy is $\mhi = (1.75 \pm 0.18) \times 10^9 \; \msun$, while the velocity 
spread between 20\% points is $\dv \approx 165$~km~s$^{-1}$ \citep[see also][]{haynes11}. No 
spectroscopically confirmed galaxy is known at 
or near the DLA redshift. The nearest galaxy, located at RA=142.50583$^\circ$, Dec=28.81740$^\circ$ (at an 
impact parameter of $\approx 7.9''$, i.e. $\approx 3.6$~kpc, to the quasar line-of-sight) is clearly 
detected in Sloan Digital Sky Survey (SDSS) images, with a photometric redshift of 
$z_{\rm phot} = 0.0207$ \citep{brescia14}, consistent with the DLA redshift. Applying the 
{\sc kcorrect} software \citep{blanton07} to the SDSS photometry of this galaxy yields a
stellar mass of $\approx 6.9^{+4.6}_{-0.5} \times 10^7 \; \msun$.

{\bf J0951+3307} $\bm{(z = 0.0054)}$:~The $z=0.0054$ DLA towards J0951+3307 has \nhi~$=(1.0 \pm 0.25) \times 10^{21}$~\cm, the highest of our sample.
The Arecibo \hi\ spectrum yields a low \hi\ mass for this DLA, $\mhi=(2.37 \pm 0.24) \times 10^8 \; 
\msun$, consistent with a dwarf galaxy. However, the \hii\ line is quite wide, with 
$\dv \approx 130$~km~s$^{-1}$. We measure log[N(S{\sc ii})/\cm]~$=(15.16 \pm 0.09)$, implying 
[M/H]~$=-0.99 \pm 0.14$. A well-known galaxy, UGC\,5282 (at $ z= 0.0052$ and at an impact parameter
of $\approx 11.2''$, i.e. $\approx 1.2$~kpc), is the likely DLA host; this has been earlier detected 
in \hii\ emission by \citet{schneider90}, with an integrated \hii\ line flux density of $(2.25 \pm 0.5)$~Jy~\kms, 
consistent within the errors with our estimate of $(1.83 \pm 0.01)$~Jy~\kms. \citet{ann15} used the SDSS 
photometry of UGC\,5282 to infer a stellar mass of $\approx 10^{7.6} \; \msun$, after correcting for the 
spatial extent of the galaxy. We obtain a gas-to-stars ratio of $M_{\rm gas}/M_{\rm stars} \approx 6$, 
not atypical for dwarf galaxies \citep[e.g.][]{begum08}.

{\bf J1415+1634} $\bm{(z = 0.0077)}$:~The $z = 0.0077$ sub-DLA towards J1415+1634 has \nhi~$=(5.0 \pm 1.2) \times 10^{19}$~\cm. The Arecibo \hii\ spectrum yields 
an \hi\ mass of $(9.65 \pm 0.97) \times 10^{8} \; \msun$, with $\dv \approx 100$~\kms. 
We obtain log[N(O{\sc i})/\cm]~$=14.48 \pm 0.04$ from the HST-COS spectrum, yielding [M/H]~$=-1.91 \pm 0.11$. 
This is a lower limit to [M/H], as ionization corrections may be significant 
for this sub-DLA. The host galaxy is likely to be UGC\,9126 (at $z = 0.007576$),
at an impact parameter of $\approx 80.1''$, i.e. $\approx 12.7$~kpc, to the quasar sightline.
\citet{wong06} estimate its integrated \hii\ line flux density to be $\approx 5.2$~Jy~\kms, 
$\approx 1.5$ times larger than our estimate for the absorber host. This is likely to be due to 
the relatively large impact parameter; note that the Arecibo primary beam  has an FWHM of 
$\approx 3.4'$. Finally, the stellar mass of UGC\,9126 has been
estimated to be quite low, $\approx (10^7 - 10^{7.42}) \; \msun$ by \citet{chang15}, via a fit to its
optical and mid-infrared spectral energy distribution. With $M_{\rm gas}/M_{\rm stars} \approx 30-100$, 
the host of the $z=0.0077$ sub-DLA appears to be an extremely gas-rich galaxy.

{\bf J1512+0128} $\bm{(z = 0.0295)}$:~We obtain \nhi~$=(2.5 \pm 0.6) \times 10^{20}$~\cm\
for the $z = 0.0295$ DLA towards J1512+0128, and $\mhi=(5.15 \pm 0.52) \times 10^9 \; \msun$, 
the highest in our sample. The \hii\ velocity spread is $\dv \approx 230$~\kms, consistent with that 
expected from a massive galaxy. We derive a metallicity of [M/H]~$=-1.29 \pm 0.13$ from the Si{\sc ii} 
line detected in the HST-COS spectrum. The absorber appears to be part of a galaxy group: six SDSS 
galaxies are present within $\approx 3'$ of the quasar line-of-sight and with redshifts within $0.005$ 
of the absorber redshift. The nearest identified galaxies with confirmed spectroscopic redshifts 
are part of a triple system (UZC-CG~236), with an impact parameter of $\approx 0.97'$ (i.e. 
$\approx 35$~kpc) to the quasar line-of-sight. However, it is possible that the absorber might 
arise in a fainter galaxy with a lower impact parameter.

{\bf J1553+3548} $\bm{(z = 0.0829)}$:~This system is also a sub-DLA, with 
\nhi~$=(5.0 \pm 1.2) \times 10^{19}$~\cm. We obtain $\mhi = (3.93 \pm 0.61) \times 10^9 \; \msun$ 
and $\dv \approx 270$~\kms\ from the \hii\ spectrum. The Si{\sc ii} line detected in the HST-COS 
spectrum yields [M/H]~$=-1.09 \pm 0.11$, before ionization corrections. \citet{battisti12} estimate 
the correction for this ion to lie in the range $[-0.38,-0.14]$; this would imply [M/H]~$=-1.35 \pm 0.16$. 
We also detect C{\sc ii}*$\lambda$1335 absorption in this absorber, yielding 
log[N(C{\sc ii}*)/\cm]=$13.41 \pm 0.14$. 
A galaxy at low impact parameter ($\approx 8.8''$) is seen in the SDSS imaging. We used the 
Low Resolution Imaging Spectrograph on the Keck telescope to 
obtain a spectrum of this galaxy (which will be discussed elsewhere) and measured $z=0.0827$ from the 
H$\beta$ and H$\alpha$ lines, indicating that this object is likely to be the DLA host.
We infer a stellar mass of $23.4^{+34.6}_{-3.9} \times 10^7 \; \msun$, applying {\sc kcorrect} to the SDSS photometry
of the galaxy.

{\bf J1619+3342} $\bm{(z = 0.0963)}$:~We were unable to detect \hii\ emission from the 
$z = 0.0963$ DLA towards J1619+3342, obtaining the $3\sigma$ upper limit $\mhi \leq 1.9 \times 10^9 \; \msun$. 
We obtain \nhi~$=(4.0 \pm 1.4) \times 10^{20}$~\cm\ from the HST-COS 
spectrum, and a metallicity of [M/H]=$-1.09 \pm 0.16$ from the S{\sc ii} line \citep[see also][]{meiring11,battisti12}. 
There are no galaxies at the DLA redshift in the SDSS spectroscopic catalog and no obvious candidate hosts close 
to the quasar line-of-sight. This system, the only absorber of our sample that does not have an 
\hii\ emission detection, also has no candidate optical counterpart, suggesting that it is likely to 
be an optically faint, low-mass galaxy.


\section{Discussion and Summary}
\label{sec:discuss}

We emphasize at the outset that our six absorbers are {\it not} an unbiased sample: J0951+3307 
and J1415+1634 were targeted with the HST due to the presence of a low-mass, gas-rich galaxy 
within 15~kpc of the quasar sightline, while J1512+0128 is part of the GASS sample \citep{borthakur15}. 
Caution must hence be taken when extending our results to general DLA samples.

Our observations have yielded \hi\ masses or \hi\ mass limits for six new DLAs and sub-DLAs.
There are now nine such absorbers with estimates of the atomic gas mass (see Table~\ref{tab:data2}). 
Five systems are DLAs, with $\nhi = (2.5-10) \times 10^{20}$~\cm, while four are sub-DLAs, with 
$\nhi = (2-7) \times 10^{19}$~\cm. Table~\ref{tab:data2} also lists the derived \nhi\ values, 
metallicities, stellar masses, the \hii\ (or CO) velocity spread between 20\% points ($\dv$), 
the velocity spread of low-ionization metal absorption lines ($\ddv$), and the gas fraction 
$f_{\rm HI} = M_{\rm HI}/M_*$ of the above nine systems. 

The \hi\ masses of the nine absorbers of our sample lie in the range $3.3 \times 10^7 - 5.2 \times 
10^9 \; \msun$, significantly lower, in all cases, than the knee of the local \hi\ mass 
function \citep[M*$_{\rm HI} \approx 10^{10} \; \msun$; e.g.][]{martin10}. 
This is consistent with expectations for the relatively low \hi\ column densities of most 
of the nine absorbers \citep[e.g.][]{zwaan05}. While the low \hi\ mass of the high-$\nhi$ 
($\approx 10^{21}$~\cm) DLA towards J0951+3307 may appear surprising, this sightline was 
selected for the presence of a low-mass, gas-rich dwarf close to the quasar sightline. Overall, 
it appears that massive galaxies do not dominate the cross-section for damped 
or sub-damped absorption at low redshifts \citep[see also ][]{zwaan05}.

While the angular resolutions of our Arecibo data ($\approx 3.4'$, i.e. $\approx 23-370$~kpc 
at the absorber redshift) are too coarse to allow a direct identification of the absorber 
host galaxies, we have used SDSS spectroscopy and photometry to identify candidate host galaxies 
for five of our six absorbers, and to estimate the stellar mass for four systems. Including systems 
from the literature, five of the six absorbers with stellar mass 
estimates have very low stellar masses ($\lesssim 2 \times 10^8 \msun$) and high gas fractions 
($f_{\rm HI} \approx 5-100$), amongst the highest gas fractions of galaxies in the local Universe 
\citep[e.g.][]{huang12}. Only two of these systems were targetted due to their gas richness. 
Further, even the sixth absorber (at $z=0.101$ towards PKS0439$-$433) has a high H$_2$ 
mass, M$_{\rm H_2} = 4.2 \times 10^9 \, \msun$ \citep{neeleman16b}, yielding a gas fraction (including 
molecular gas) of $ 0.42 \lesssim f_{\rm HI+H_2} \lesssim 0.7$, quite high for late-type disk galaxies.
The metallicities of most absorbers are relatively low, with seven of the nine systems having
[M/H]~$\lesssim -1$; this is surprising for objects in the nearby Universe \citep[although 
consistent with metallicity evolution in DLAs; ][]{rafelski12}. 
Most of the absorbers of the sample thus appear to arise in gas-rich galaxies, with low star 
formation activity. However, the \hii\ velocity spreads are too large to be explained by an origin 
in individual dwarf galaxies. It is possible that the large velocity widths arise from \hii\ 
emission from multiple faint galaxies in the relatively large Arecibo beam; we are now 
investigating this with interferometric \hii\ studies.

In conclusion, we have used a combination of HST-COS FUV spectroscopy, Arecibo \hii\ 
emission spectroscopy and SDSS photometry to estimate or constrain the \hi\ column density, 
the metallicity, the atomic gas mass, and the stellar mass for a sample of six DLAs and 
sub-DLAs at $z \lesssim 0.1$. We obtain \hi\ masses $\approx (0.2 - 5) \times 10^9 \msun$, stellar masses 
$\lesssim 5 \times 10^8 ~\msun$, low metallicities, $\lesssim 0.1$~solar, and high gas fractions, 
$f_{\rm HI} \equiv \mhi/{M_*} \gtrsim 5$ (amongst the highest in the nearby Universe), for the absorbers of 
our sample. The large velocity spreads of the \hii\ and CO emission lines ($\dv \approx 100-290$~\kms), 
the high gas fractions ($f_{\rm HI} \approx 5-100$), and the low metallicities and stellar masses suggest 
that the absorbers are gas-rich galaxies with a low star formation efficiency.

\section*{Acknowledgments}

It is a pleasure to thank Chris Salter and Phil Perrilat for much help with the Arecibo
observations. NK acknowledges support from the Department of Science and Technology via 
a Swarnajayanti Fellowship (DST/SJF/PSA-01/2012-13). The Arecibo Observatory is operated 
by SRI International under a cooperative agreement with the National Science Foundation 
(AST-1100968), and in alliance with Ana G. M\'{e}ndez-Universidad Metropolitana, and the 
Universities Space Research Association. 

\label{lastpage}

\bibliographystyle{mn2e}
\bibliography{ms}

\end{document}